\documentclass{iaus}
\usepackage{graphicx}

\title[short title of paper] 
{ Exploring variations in the fundamental constants with
  ELTs: The CODEX spectrograph on OWL}

\author[short author list]   
{Paolo Molaro$^{1,2}$,
 Michael T.~Murphy$^3$ \break \and Sergei A.~Levshakov$^4$ for the CODEX team \thanks{L. Pasquini, H. Dekker,
S. Cristiani, F. Pepe, M. Haehnelt, G. Avila, B. Delabre, , S. D'Odorico, J.
Liske, P. Shaver, P. Bonifacio, S. Borgani, V. D'Odorico, E. Vanzella, M.
Dessauges-Zavadsky C. Lovis, M. Mayor, M. Viel, F. Bouchy, A. Grazian, L.
Moscardini, T. Wicklind, S. Zucker.}}
\affiliation{$^1$ Osservatorio Astronomico di Trieste, Via G.~B.~Tiepolo 11,
  34131 Trieste, Italy 
\\[\affilskip]
$^2$Observatoire de Paris 61, avenue de l'Observatoire, 75014 Paris, France
\\[\affilskip]
$^3$  Institute of Astronomy, University of Cambridge, Madingley Road,
Cambridge CB3 0HA, UK  
\\[\affilskip]
$^4$   Department of Theoretical Astrophysics,
Ioffe Physico-Technical Institute, 194021 St.Petersburg, Russia 
}

\pubyear{2006}
\volume{232}  
\pagerange{119--126}
\date{?? and in revised form ??}
\jname{The Scientific Requirements for Extremely Large Telescopes}
\editors{P., Whitelock, B. Leibundgut  \& M. Dennefeld, eds.}
\begin{document}

\maketitle

\begin{abstract}
Cosmological variations in the fine structure constant, $\alpha$, can
be probed through precise velocity measurements of metallic absorption
lines from intervening gas clouds seen in spectra of distant quasars.
Data from the Keck/HIRES instrument support a variation in $\alpha$ of
6 parts per million. Such a variation would have profound
implications, possibly providing a window into the extra spatial
dimensions required by unified theories such as string/M-theory.
However, recent results from VLT/UVES suggest no variation in
$\alpha$.   The COsmic
Dynamics EXperiment (CODEX) spectrograph currently being designed for
the ESO OWL telescope (Pasquini et al 2005) with a  resolution high enough
to properly resolve
even the narrowest of metallic absorption lines, $R>150\,000$, will achieve
a 2-to-3 order-of-magnitude
precision increase in $\Delta\alpha/\alpha$. This will rival the
precision available from the Oklo natural fission reactor and upcoming
satellite-borne atomic clock experiments. Given the vital constraints
on fundamental physics possible, the ELT community must consider such
a high-resolution optical spectrograph like CODEX.  
\end{abstract}

\firstsection 
\section{Introduction}

 Experimental
exploration of possible space-time variations in fundamental constants
 has focused on the dimensionless fine-structure
constant, {$\alpha =
  \frac{e^2}{\hbar c}$}, {$ = 1/137.035 999 11(46)$}. In quantum
electrodynamics, $\alpha$ plays the role of a
coupling constant, representing the strength of the interaction
between electrons and photons. It cannot be predicted by 
theory, and  it is
one of the twenty-odd "external parameters" in the Standard Model of
particle physics.

Many modern theories predict variations in various
fundamental constants  (e.g.~see review in \cite{Uzan03}).   
String theory    implies
supersymmetry and predicts six undiscovered dimensions of space.
 In  higher dimensional theories the
constants' values are defined in the full dimensional space, and any
evolution either in time or space in their  sizes  would
lead to
dynamical constants in the 4D space. Varying $\alpha$  is obtained by
arbitrarily coupling photon to scalar fields.
Scalar fields provide negative pressure and may drive the 
cosmological acceleration (Bento et al 2004, Fujii
2005).  If the scalar fields are
extremely light they could produce variations of constants only on
cosmological time scales.  The functional dependence of the gauge-coupling
constants on
cosmological time is not known and even oscillations might be possible
during the course of the cosmological evolution ( \cite{Marciano84}).
  In this regard astronomical observations are the only way to
test such predictions at different space-time coordinates.


\begin{figure}[h]
\begin{center}
\includegraphics[width=.65\textwidth]{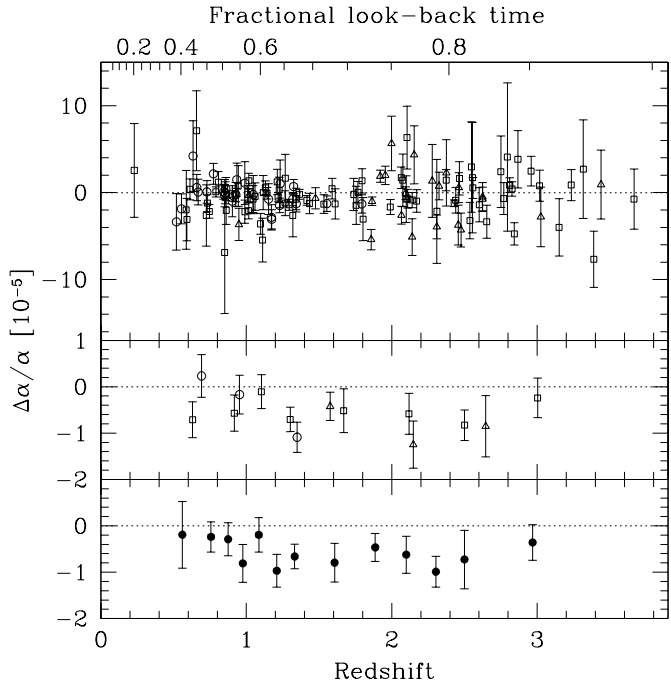}
\end{center}
\caption[]{Murphy et al 2004. The  variation in $\alpha$ is found at  5
$\sigma$ confidence level.}
\label{eps1}
\end{figure}

\section{Measures and constraints on variability}

Big Bang nucleosynthesis  and the cosmic microwave background power
 spectrum constrains $\Delta \alpha / \alpha  = (\alpha_z - \alpha_0)/\alpha_0 \le 10^{-2}$
 at z$\approx 10^{10}$ and at z$\approx 1100$, respectively. Atomic clocks
in laboratory experiments restrict the
 time-dependence of $\alpha$ 
   at 
the level of {$ (\dot{\alpha}/ \alpha)_{t_0}  \approx   10^{-15}$
yr$^{-1}$}, or  {$|\Delta \alpha/\alpha|
<  10^{-5}$},   for  a  time  of
$t \sim 10^{10}$ yr\, assuming  $\alpha_z$ changing  linearly with time. 
The ACES (Atomic Clock Ensemble in Space) project   forseen to fly on the
International Space Station in 2007, will  operate a cold atom 
clock in microgravity to test General relativity and 
search for a possible drift of the fine structure constant reaching the 
10$^{-17}$ yr$^{-1}$ stability range with a gain of two orders 
of magnitude with respect 
to present laboratory constraints.
Meteoritic data on  the radioactive $\beta$-decay
of  $^{187}$Re   place a bound  around
  $\Delta \alpha / \alpha  \le 10^{-7}$ (Olive et al 2004), and a recent
analysis 
of the isotopic abundance in the
Oklo samples  provides a hint for a variation although at a very low level: { 
$\Delta \alpha / \alpha \ge 4.5 \cdot 10^{-8}$}  (Lamoreaux \& Torgerson,
2004).

\begin{figure}[h]
\begin{center}
\includegraphics[width=.65\textwidth]{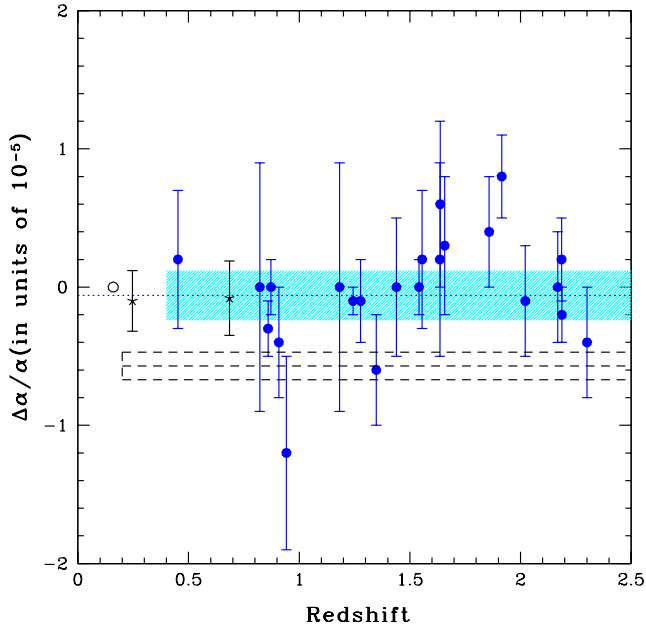}
\end{center}
\caption[]{Chand et al. (2004)   measures in  23  absorption systems with
 VLT/UVES. 
The Keck results are also shown with the dashed line rectangle}
\label{eps1}
\end{figure}
 
High resolution spectroscopy of  absorption systems lying
along the lines-of-sight to background QSOs followed two approaches. One
focused on the alkali-doublets (AD) 
 since the comparison between AD separations seen in absorption systems
with those measured in the laboratory provides a simple probe
of  variation of $\alpha$.   The  current  more stringent  constraints
come from
analysis of  Si IV absorption systems in 15 systems with redshift 1.6$<
z_{abs} < $ 3: {$\Delta \alpha / \alpha$ = (1.5 $\pm$ 4.3) Ã 10$^{-6}$}
(Chand et al.
2005). A second approach, the multi multiplet (MM)  compare the line
shifts of the species particularly sensitive to a change 
in $\alpha$ such as  Cr, Fe, Ni and Zn 
to one with a comparatively minor sensitivity, such as  Mg, Si and Al 
which is referred to as an anchor  
 (Dzuba, Flambaum \& Webb
1999 and Webb et al. 1999). The sensitivity  of each line
transition  on a change in $\alpha$ is expressed  by a  coefficient $q$
(Dzuba et al. 1999, 2002). 

Applied to Keck/HIRES QSO absorption
spectra the MM method has yielded  the first  evidence for a
varying $\alpha$ by Webb et al. (1999) becoming
stronger with successively larger samples. 
The most
recent  value { $ \Delta \alpha / \alpha = (-5.7 \pm 1.1) Ã
10^{-6}$} comes from the analysis of 143 absorption systems
over the range 0.2 $< z_{abs} < $4.2 and it is shown in Fig 1 (Murphy et
al 2004).  The deduced variation in $\alpha$ at about 5 $\sigma$
CL and the implications are extraordinary. 
However,  the result has not been confirmed by other authors.  Chand et
al. (2004)   have analyzed 23 Mg,Fe absorption systems in higher
signal-to-noise ratio (S/N) spectra from a different telescope and
spectrograph, the VLT/UVES, claiming a  null result over
the range 0.4 $< z_{abs} <$ 2.3 { $ \Delta \alpha / \alpha = (-0.6 \pm
0.6) Ã 10^{-6}$}, which is shown in Fig 2. 
A  variant of this methodology   makes use only of  pairs of FeII lines
which differ  significantly in their   $q$ values, in order to minimize
 potential systematic errors. This  has been applied in  a couple of
systems,  at 
$z_{abs}$ = 1.839 in the spectrum of Q1101--264   and 
  at  $z_{abs}$ = 1.15   in the spectrum of 
HE0515--4414  providing  $\Delta\alpha / \alpha  = (0.4\pm1.5_{\rm
stat})\times10^{-6}$ 
and  {$\Delta\alpha / \alpha  = (-0.07\pm0.84_{\rm stat})\times10^{-6}$,
respectively  (Levshakov et al 2005, 2006a).
The discrepancy between the VLT/UVES and Keck/HIRES results is yet to
be resolved and demands for a high resolution spectrograph at an ELT for
significant progress.

It is also interesting to note  that  Fujii (2005) has  been able to fit
both  the Chand et al and  the Levshakov et al null results 
by means of an oscillatory behaviour for a varying $\alpha$. A behaviour
which can even 
 traced by eye in the original data of Chand et al shown in Fig 2. To
probe such an intriguing possibility  very high precision measurements
 at individual redshifts are required. 
Moreover,  the MM measures which use {\rm Mg} as an anchor relay on terrestrial
relative composition of the {\rm Mg} isotopes. Since the frequency shifts of
$\Delta\alpha / \alpha$  are of the order of magnitude of the typical
isotope shifts a departure from these values can produce different
 results for the measures which take {\rm Mg}  as an anchor.  Sub-solar
$^{25,26}{\rm Mg}/^{24}{\rm Mg}$ ratios would make a variation of $\alpha$ 
even more significant. 
  Suggestion that this my be the case comes indirectly from a recent  
upper limit in the $^{13}{\rm C}$ abundance ($^{12}{\rm C}/^{13}{\rm C} > 200$, 1$\sigma$),
in  the system at  $z_{abs}$ = 1.15  in the spectrum of 
HE0515--4414 (Levshakov et al 2006b). Since both $^{25,26}{\rm Mg}$ and 
 $^{13}{\rm C}$ are produced in the Hot Bottom Burning stage  of AGBs, a   low  $^{13}{\rm C}$
 imply possibly low  $^{25,26}{\rm Mg}$. In the case of Chand et al data set the
relaxation of this assumption  would have produced
 a $\Delta\alpha / \alpha  = (-3.6\pm0.6_{\rm stat})\times10^{-6}$. This
well illustrate that the case for a variability require a 
better understanding of the isotopic  evolution of the absorption clouds,
a problem which can be addressed
only with a spectrograph of very high resolution able to separate the
isotopic lines such as the one foreseen for CODEX.

\begin{figure}[h]
\begin{center}
\includegraphics[width=.65\textwidth]{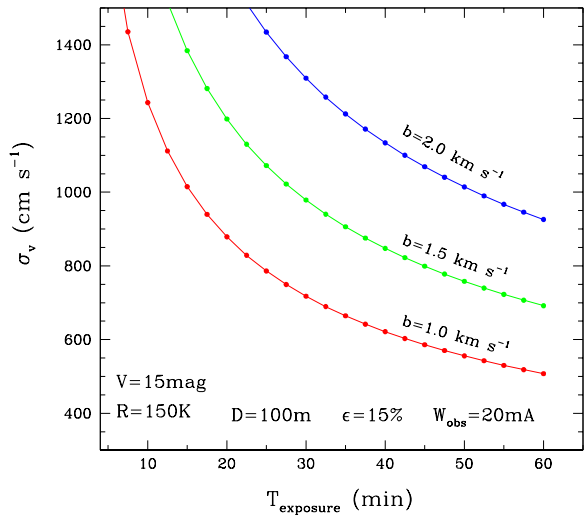}
\end{center}
\caption[]{Wavelength precision measurement of a single metal line with
 broadening of 1, 1.5 and 2 km s$^{-1}$, with a spectrograph 
with R= 150000 at a 100 m telescope. See text for details.}
\label{eps1}
\end{figure}

\section{With CODEX @ OWL}

A   measurement of   $\Delta\alpha/\alpha$ is essentially a measurement of the
wavelength for a pair or more  lines with different
 sensitivity coefficients. Therefore the  accuracy of a  
$\Delta\alpha/\alpha$ measurement is ultimately determined by the
precision  with which a line position can be determined in the spectrum.
With current spectrographs with R$= \lambda/\delta\lambda \approx$ 4$\cdot
10^{4}$ 
 the observed line positions can be  known
with the accuracy of  about $\sigma_\lambda \approx$ 1 m\AA\ ( or $\Delta
v = 60$ m s$^{-1}$ at 5000 \AA), then
$\tilde{\sigma}_{\Delta\alpha/\alpha} \simeq 10^{-5}$  for a 
typical pair. This value is normally  further improved to 
 reach one part per million when more transitions and/or  more systems are
available.
Any improvement with respect to this figure is related to  the possibility
to measure line position more accurately. This  can be achieved
  with an increase in the resolving power of the spectrograph up to the
point in which  the narrowest 
lines formed in intervening physical clouds are resolved, and  with an
increase of the signal-to-noise ratio in the spectrum (Bohlin et al. 1983). 
The metal  lines which are  observed in the QSO absorption  systems have 
intrinsic widths  of  few km s$^{-1}$, 
rarely of less than 1 km s$^{-1}$. The CODEX spectrograph with a 
resolving power R$\approx$ 150 000 match well these requirements.

In Fig 3  the precision in velocity in the measurement of the
absorption line as a function of  exposure time at the OWL telescope is shown.
We used the Bohlin et al formula for  an absorption line  with an equivalent
width  of 0.02 $\AA$  and $ b$ values of 1, 1.5 and 2  km  s$^{-1}$. The spectrograph is
the CODEX one as it is described in Pasquini et al (2006) and a global 
 efficiency of 15$\%$ is assumed.
 A  precision of 5-10 m  s$^{-1}$ 
is obtained in a relatively short time of 1 hours. This  would imply a
precision in the  determination of $\alpha$ 
 of the order of $\tilde{\sigma}_{\Delta\alpha/\alpha} \simeq 10^{-7}$   
 for a single pair of lines. The precision  can
then  further increased by extending the exposure time or by statistics using other 
lines belonging to  the same system 
or more systems with similar redshift. The overall final  precision will rival, and
probably surpass,  the one
 available from the Oklo natural fission reactor and upcoming
satellite-borne atomic clock experiments. Given the deep insights on
several aspects of fundamental physics 
and cosmology, the ELT community must consider such
a high-resolution optical spectrograph like CODEX.

\begin{acknowledgments}
We  acknowledge the whole  CODEX team
\end{acknowledgments}

\end{document}